# Cg in Two Pages

Mark J. Kilgard
NVIDIA Corporation
Austin, Texas
January 16, 2003

## 1. Cg by Example

Cg is a language for programming GPUs. Cg programs look a lot like C programs. Here is a Cg vertex program:

```
void simpleTransform(float4 objectPosition : POSITION,
                     float4 color          : COLOR,
                     float4 decalCoord     : TEXCOORD0,
                     float4 lightMapCoord  : TEXCOORD1,
                 out float4 clipPosition   : POSITION,
                 out float4 oColor         : COLOR,
                 out float4 oDecalCoord    : TEXCOORD0,
                 out float4 oLightMapCoord : TEXCOORD1,
             uniform float brightness,
             uniform float4x4 modelViewProjection)
{
  clipPosition = mul(modelViewProjection, objectPosition);
  oColor = brightness * color;
  oDecalCoord = decalCoord;
  oLightMapCoord = lightMapCoord;
}
```

### 1.1 Vertex Program Explanation

The program transforms an object-space position for a vertex by a 4x4 matrix containing the concatenation of the modeling, viewing, and projection transforms. The resulting vector is output as the clip-space position of the vertex. The per-vertex color is scaled by a floating-point parameter prior to output. Also, two texture coordinate sets are passed through unperturbed.

Cg supports scalar data types such as `float` but also has first-class support for vector data types. `float4` represents a vector of four floats. `float4x4` represents a matrix. `mul` is a standard library routine that performs matrix by vector multiplication. Cg provides function overloading like C++; `mul` is an overloaded function so it can be used to multiply all combinations of vectors and matrices.

Cg provides the same operators as C. Unlike C however, Cg operators accept and return vectors as well as scalars. For example, the scalar, `brightness`, scales the vector, `color`, as you would expect.

In Cg, declaring a parameter with the `uniform` modifier indicates that its value is initialized by an external source that will not vary over a given batch of vertices. In this respect, the `uniform` modifier in Cg is different from the `uniform` modifier in RenderMan but used in similar contexts. In practice, the external source is some OpenGL or Direct3D state that your application takes care to load appropriately. For example, your application must supply the `modelViewProjection` matrix and the `brightness` scalar. The Cg runtime library provides an API for loading your application state into the appropriate API state required by the compiled program.

The `POSITION`, `COLOR`, `TEXCOORD0`, and `TEXCOORD1` identifiers following the `objectPosition`, `color`, `decalCoord`, and `lightMapCoord` parameters are input semantics. They indicate how their parameters are initialized by per-vertex varying data. In OpenGL, `glVertex` commands feed the `POSITION` input semantic; `glColor` commands feed the `COLOR` semantic; `glMultiTexCoord` commands feed the `TEXCOORD`n semantics.

The `out` modifier indicates that `clipPosition`, `oColor`, `oDecalCoord`, and `oLightMapCoord` parameters are output by the program. The semantics that follow these parameters are therefore output semantics. The respective semantics indicate the program outputs a transformed clip-space position and a scaled color. Also, two sets of texture coordinates are passed through. The resulting vertex is feed to primitive assembly to eventually generate a primitive for rasterization.

Compiling the program requires the program source code, the name of the entry function to compile (`simpleTransform`), and a profile name (`vs_1_1`).

The Cg compiler can then compile the above Cg program into the following DirectX 8 vertex shader:

```
vs.1.1
mov oT0, v7
mov oT1, v8
dp4 oPos.x, c1, v0
dp4 oPos.y, c2, v0
dp4 oPos.z, c3, v0
dp4 oPos.w, c4, v0
mul oD0, c0.x, v5
```

The profile indicates for what API execution environment the program should be compiled. This same program can be compiled for the DirectX 9 vertex shader profile (`vs_2_0`), the multi-vendor OpenGL vertex program extension (`arbvp1`), or NVIDIA-proprietary OpenGL extensions (`vp20` & `vp30`).

The process of compiling Cg programs can take place during the initialization of your application using Cg. The Cg runtime contains the Cg compiler as well as API-dependent routines that greatly simplify the process of configuring your compiled program for use with either OpenGL or Direct3D.

### 1.2 Fragment Program Explanation

In addition to writing programs to process vertices, you can write programs to process fragments. Here is a Cg fragment program:

```
float4 brightLightMapDecal(float4 color        : COLOR,
                           float4 decalCoord   : TEXCOORD0,
                           float4 lightMapCoord : TEXCOORD1,
                    uniform sampler2D decal,
                    uniform sampler2D lightMap) : COLOR
{
  float4 d = tex2Dproj(decal, decalCoord);
  float4 lm = tex2Dproj(lightMap, lightMapCoord);
  return 2.0 * color * d * lm;
}
```

The input parameters correspond to the interpolated color and two texture coordinate sets as designated by their input semantics.

The `sampler2D` type corresponds to a 2D texture unit. The Cg standard library routine `tex2Dproj` performs a projective 2D texture lookup. The two `tex2Dproj` calls sample a decal and light map texture and assign the result to the local variables, `d` and `lm`, respectively.

The program multiplies the two textures results, the interpolated color, and the constant 2.0 together and returns this RGBA color. The program returns a `float4` and the semantic for the return value is `COLOR`, the final color of the fragment.

The Cg compiler generates the following code for `brightLightMapDecal` when compiled with the `arbfp1` multi-vendor OpenGL fragment profile:

```
!!ARBfp1.0
PARAM c0 = {2, 2, 2, 2}; TEMP R0; TEMP R1; TEMP R2;
TXP R0, fragment.texcoord[0], texture[0], 2D;
TXP R1, fragment.texcoord[1], texture[1], 2D;
MUL R2, c0.x, fragment.color.primary;
MUL R0, R2, R0;
MUL result.color, R0, R1;
END
```

This same program also compiles for the DirectX 8 and 9 profiles (`ps_1_3` & `ps_2_x`) and NVIDIA-proprietary OpenGL extensions (`fp20` & `fp30`).

## 2. Other Cg Functionality

### 2.1 Features from C

Cg provides structures and arrays, including multi-dimensional arrays. Cg provides all of C's arithmetic operators (`+`, `*`, `/`, etc.). Cg provides a `boolean` type and boolean and relational operators (`||`, `&&`, `!`, etc.). Cg provides increment/decrement (`++`/`--`) operators, the conditional expression operator (`?:`), assignment expressions (`+=`, etc.), and even the C comma operator.

Cg provides user-defined functions (in addition to pre-defined standard library functions), but recursive functions are not allowed. Cg provides a subset of C's control flow constructs (`do`, `while`, `for`, `if`, `break`, `continue`); other constructs such as `goto` and `switch` are not supported in current the current Cg implementation but the necessary keywords are reserved.

Like C, Cg does not mandate the precision and range of its data types. In practice, the profile chosen for compilation determines the concrete representation for each data type. `float`, `half`, and `double` are meant to represent continuous values, ideally in floating-point, but this can depend on the profile. `half` is intended for a 16-bit half-precision floating-point data type. (NVIDIA's CineFX architecture provides such a data type.) `int` is an integer data type, usually used for looping and indexing. `fixed` is an additional data type intended to represent a fixed-point continuous data type that may not be floating-point.

Cg provides `#include`, `#define`, `#ifdef`, etc. matching the C preprocessor. Cg supports C and C++ comments.

### 2.2 Additional Features Not in C

Cg provides built-in constructors (similar to C++ but not user-defined) for vector data types:

```
float4 vec1 = float4(4.0, -2.0, 5.0, 3.0);
```

Swizzling is a way of rearranging components of vector values and constructing shorter or longer vectors. Example:

```
float2 vec2 = vec1.yx;      // vec2 = (-2.0, 4.0)
float scalar = vec1.w;      // scalar = 3.0
float3 vec3 = scalar.xxx;   // vec3 = (3.0, 3.0, 3.0)
```

More complicated swizzling syntax is available for matrices. Vector and matrix elements can also be accessed with standard array indexing syntax as well.

Write masking restricts vector assignments to indicated components. Example:

```
vec1.xw = vec3;   // vec1 = (3.0, -2.0, 5.0, 3.0)
```

Use either `.xyzw` or `.rgba` suffixes swizzling and write masking.

The Cg standard library includes a large set of built-in functions for mathematics (`abs`, `dot`, `log2`, `reflect`, `rsqrt`, etc.) and texture access (`texCUBE`, `tex3Dproj`, etc.). The standard library makes extensive use of function overloading (similar to C++) to support different vector lengths and data types. There is no need to use `#include` to obtain prototypes for standard library routines as in C; Cg standard library routines are automatically prototyped.

In addition to the `out` modifier for *call-by-result* parameter passing, the `inout` modifier treats a parameter as both a *call-by-value* input parameter and a *call-by-result* output parameter.

The `discard` keyword is similar to `return` but aborts the processing without returning a transformed fragment.

### 2.3 Features Not Supported

Cg has no support currently for pointers or bitwise operations (however, the necessary C operators and keywords are reserved for this purpose). Cg does not (currently) support unions and function variables.

Cg lacks C++ features for "programming in the large" such as classes, templates, operator overloading, exception handling, and namespaces.

The Cg standard library lacks routines for functionality such as string processing, file input/output, and memory allocation, which is beyond the specialized scope of Cg.

However, Cg reserves all C and C++ keywords so that features from these languages could be incorporated into future implementations of Cg as warranted.

## 3. Profile Dependencies

When you compile a C or C++ program, you expect it to compile without regard to how big (within reason) the program is or what the program does. With Cg, a syntactically and semantically correct program may still not compile due to limitations of the profile for which you are compiling the program.

For example, it is currently an error to access a texture when compiling with a vertex profile. Future vertex profiles may well allow texture accesses, but existing vertex profiles do not. Other errors are more inherent. For example, a fragment profile should not output a parameter with a `TEXCOORD0` semantic. Other errors may be due to exceeding a capacity limit of current GPUs such as the maximum number of instructions or the number of texture units available.

Understand that these profile dependent errors do not reflect limitations of the Cg language, but rather limitations of the current implementation of Cg or the underlying hardware limitations of your target GPU.

## 4. Compatibility and Portability

NVIDIA's Cg implementation and Microsoft's High Level Shading Language (HLSL) both implement the same language. While HLSL is tightly integrated to DirectX 9 and the Windows operating systems, Cg provides support for multiple APIs (OpenGL, DirectX 8, and DirectX 9) and multiple operating systems (Windows, Linux, and Mac OS X). Because Cg interfaces to multi-vendor APIs, Cg runs on GPUs from multiple vendors.

## 5. More Information

In March 2003, look for *The Cg Tutorial: The Definitive Guide to Programmable Real-Time Graphics* (ISBN 0321194969) published by Addison-Wesley.